\documentstyle[fleqn,espcrc2,epsf]{article}
\title{
$I=2$ Pion Scattering Phase Shift with Wilson Fermions
\thanks{presented by N. Ishizuka}
}
\author{
CP-PACS Collaboration : 
S.~Aoki
\address{
Institute of Physics,
University of Tsukuba,
Tsukuba, Ibaraki 305-8571, Japan
},
M.~Fukugita
\address{
Institute for Cosmic Ray Research,
University of Tokyo,
Kashiwa 277-8582, Japan
},
S.~Hashimoto
\address{
High Energy Accelerator Research Organization(KEK),
Tsukuba, Ibaraki 305-0801, Japan
},
K-I.~Ishikawa$^{\rm ~a,}$
\address{
Center for Computational Physics,
University of Tsukuba,Tsukuba, Ibaraki 305-8577, Japan
},
N.~Ishizuka$^{\rm ~a,d}$,
Y.~Iwasaki$^{\rm ~a}$,
K.~Kanaya$^{\rm ~a}$,
T.~Kaneko$^{\rm ~c}$,
Y.~Kuramashi$^{\rm ~c}$,
V.~Lesk$^{\rm ~d}$,
M.~Okawa
\address{
Department of Physics, Hiroshima University,
Higashi-Hiroshima 739-8526, Japan
},
Y.~Taniguchi$^{\rm ~a}$,
A.~Ukawa$^{\rm ~a,d}$ and
T.~Yoshi\'e$^{\rm ~a,d}$ 
}
\begin{document}
%
%
\begin{abstract}
We present results of phase shift for $I=2$ $S$-wave $\pi\pi$ system 
with the Wilson fermions in the quenched approximation.
The finite size method proposed by L\"uscher is employed, 
and calculations are carried out at $\beta=5.9$ ($a^{-1}=1.934(16)$~GeV from 
$m_\rho$) 
on
$24^3 \times 60$ , 
$32^3 \times 60$ , and
$48^3 \times 60$ lattices. 
\end{abstract}
\maketitle
%
%
\vspace{-0.2cm}
\section{ Introduction }
Lattice calculation of phase shifts is an important step for understanding 
of strong interactions beyond the hadron mass spectrum.
For scattering lengths which are the threshold values of phase shifts, 
there are already several studies in literature.
For the $I=2$ $\pi\pi$ scattering calculations have been carried out with 
the staggered~\cite{SGK,Kuramashi},  
the standard~\cite{Kuramashi,GPS,JLQCD} and 
the improved Wilson fermion actions~\cite{LZCM}.
Recent studies by JLQCD~\cite{JLQCD} and by Liu {\it et.al.}~\cite{LZCM}
carried out the calculation at several lattice cutoffs
and obtained the scattering length in the continuum limit.
For the scattering phase shift, in contrast, 
there is only one calculation for $I=2$, 
which used lattice data to estimate an effective $\pi\pi$ potential and a quantum mechanical 
analysis to calculate the phase shift from the potential~\cite{Fiebig}.

We present a lattice QCD calculation of the $I=2$ $S$-wave $\pi\pi$ phase shift
using the finite-size method~\cite{Lusher}.
We work in quenched lattice QCD employing the standard plaquette 
action for gluons at $\beta = 5.9$ and the Wilson fermion action for quarks.
The number of configurations (lattice sizes) are
$200$ ($24^3\times 60$),
$286$ ($32^3\times 60$), and
$ 56$ ($48^3\times 60$).
Quark propagators are solved with the Dirichlet boundary condition 
imposed in the time direction and the periodic boundary condition in the 
space directions. 
Quark masses are chosen to be the same as in the previous study
of quenched hadron spectroscopy by CP-PACS~\cite{CP-PACS.LHM}
({\it i.e.}, $m_\pi/m_\rho = 0.491, 0.593, 0.692, 0.752$ ) 
for all lattice sizes.
Preliminary results of the present work 
was presented at Lattice'01~\cite{CP-PACS.PHSH.old}.
%
%
\vspace{-0.2cm}
\section{ Method }
The energy eigenvalue $W_p$ of an $S$-wave $\pi\pi$ system
with momentum $\vec{p}$ and $-\vec{p}$ 
in a finite periodic box of a size $L^3$
is shifted from twice the pion energy $2\cdot E_p$ by finite-size effects.
L\"uscher derived a relation between the energy shift 
$\Delta W_p = W_p - 2\cdot E_p$
and the scattering phase shift $\delta(p)$, 
which takes the form \cite{Lusher}
\begin{equation}
\nu_n \frac{ \tan \delta (p) }{ \pi L p } = - x_p - A_n x_p^2 - B_n x_p^3 + O(x_p^4)
\label{Lusher.eq}
\ ,
\end{equation}
where
$p^2 = n \cdot (2\pi/L)^2$, ($n = 0, 1, \cdots 6$),
$x_p = \Delta W_p \cdot (2 E_p L^2) / (16\pi^2) = O(1/ L)$, 
and $\nu_n$, $A_n$, and $B_n$ are geometrical constants.
The scattering length is given by $a_0 = ( \tan \delta(p) / p )_{p \to 0}$. 

In order to obtain the energy eigenvalue $W_p$
we construct $\pi\pi$ 4-point functions 
$G_{pk}(t) = \langle 0 | \Omega_p ( t ) \Omega_k ( 0 ) | 0 \rangle$.
Here $\Omega_p( t )$ is an interpolating field for the $S$-wave $\pi\pi$ system 
at time $t$ given by 
$\Omega_p( t ) = \sum_{R}
                      \pi(  R(\vec{p}), t )
                      \pi( -R(\vec{p}), t )$ 
where $R$ is an element of the cubic group. 
In numerical calculations
we construct the source operator $\Omega_k( 0 )$
using the noisy source method with U(1) random numbers.
%
%
\begin{figure}[t]
\vspace*{-0.7cm}
\centerline{\epsfxsize=7.0cm \epsfbox{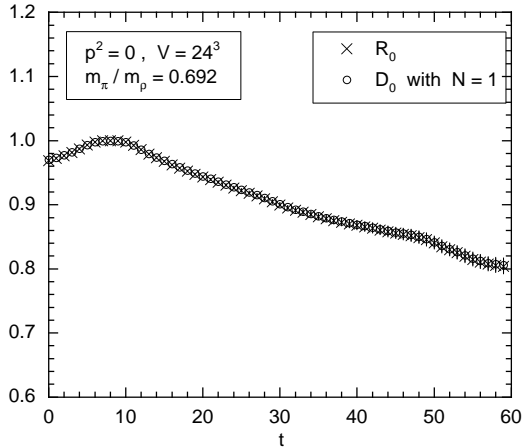}}
\vspace*{-1.0cm}
\caption{\label{Fig1.fig}
$R_p(t)$ and $D_p(t)$ for $p^2=0$.
}
\vspace*{-0.6cm}
\end{figure}
%
%

Since the $4$-point function $G_{pk}(t)$ contains many exponential terms
as pointed out by Maiani and Testa~\cite{MT-nogo},
the extraction of energy eigenvalues from $G_{pk}(t)$ is non-trivial. 
We solve this problem by the method proposed by L\"uscher and Wolf~\cite{Lusher-Wolf}.
In their method, 
one diagonalize the $M_{pk} (t_0,t)$ $\equiv$ $[ G(t_0)^{-1/2}$ $G(t) G(t_0)^{-1/2}]_{pk}$
at each $t$ where $t_0$ is fixed at some small value.
The eigenvalues are given by $\lambda_p (t) = \exp( - W_p (t-t_0) )$.

In the actual diagonalization we have to cut off the set of momenta.
Here we expect that the components of $G_{pk}(t)$ or $M_{pk}(t_0,t)$ for 
$p,k \leq q$ are dominant for the eigenvalue $\lambda_q(t)$ in the large $t$ 
region, while the components $p,k > q$ are less important.

With this expectation, we calculate the energy eigenvalue
for the ground state ($n=0$) and the first excited state ($n=1$) for $V=24^3$,
and also that of the second excited state ($n=2$) for $V=32^3$ and $V=48^3$.
The cut-off dependence is investigated by varying the number of 
momenta $N \geq n$ for $N=0,1,2,3$.
%
%
\begin{figure}[t]
\vspace*{-0.7cm}
\centerline{\epsfxsize=7.0cm \epsfbox{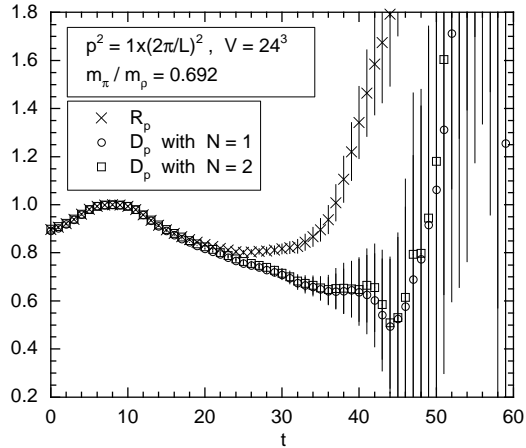}}
\vspace*{-1.0cm}
\caption{\label{Fig2.fig}
$R_p(t)$ and $D_p(t)$ for $p^2=(2\pi/L)^2$.
}
\vspace*{-0.6cm}
\end{figure}
%
%
\vspace{-0.2cm}
\section{ Results }
In order to examine the effects of diagonalization,
we calculate two ratios defined by
$R_{p}(t) \equiv G_{pp}(t) / [ G_{p}^\pi (t) ]^2$
and 
$D_{p}(t) \equiv \lambda_p (t) / [ G_p^\pi (t_0) / G_p^\pi (t) ]^2$,
where $G_{p}^\pi (t)$ is the pion propagator with momentum $p$.
If $G_{pp}(t)$ or $\lambda_p(t)$ behaves as a single exponential function, 
we can obtain the energy shift $\Delta W_p$ 
from the ratio $R_p(t)$ or $D_p(t)$ by a single exponential fit.

In Fig.~\ref{Fig1.fig} we compare the ratios for $p^2=0$. 
The two-pion source is placed at $t=8$.  
The momentum cut-off is set at $N=1$.
The signals are very clear and diagonalizations do not affect the result. 
We checked the cut-off dependence by taking $N=2$ and confirmed
that it is negligible.
In previous calculations of scattering lengths\cite{SGK,Kuramashi,GPS,JLQCD}
the ratio $R_0(t)$ was used to extract the energy shift $\Delta W_0$.
Our calculation demonstrates the reliability of these calculations.

We compare the ratios for $p^2=(2\pi/L)^2$ in Fig.~\ref{Fig2.fig}.
The momentum cutoff is set at $N=1$ and $N=2$.
In contrast to the $p^2=0$ case, the diagonalization is very effective. 
We can observe a convincing single exponential behavior only after the 
diagonalization. The cut-off dependence is negligible. 

The analysis shown here, and additional ones for $p^2=(2\pi/L)^2 \cdot 2$ 
for $V=32^3$ and $V=48^3$, 
lead us to conclude that 
the momentum cut-off should be taken $N \geq n$
for the energy shift $\Delta W_p$ ( $p^2=(2\pi/L)^2\cdot n$ ).
%
%
\begin{figure}[t]
\vspace*{-0.7cm}
\centerline{\epsfxsize=7.0cm \epsfbox{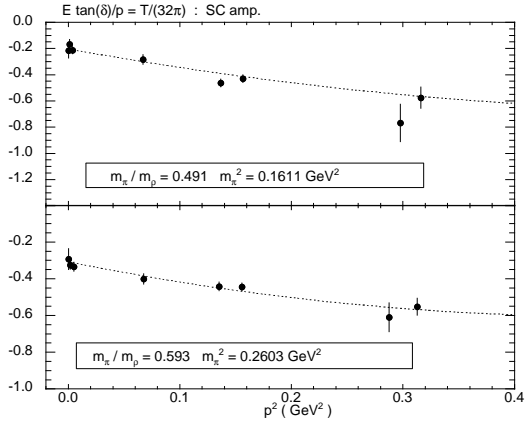}}
\vspace*{-1.0cm}
\caption{\label{Fig3.fig}
Results for scattering amplitude.
}
\vspace*{-0.8cm}
\end{figure}
%
%

In Fig.~\ref{Fig3.fig}
we plot our results for the scattering amplitude
$T/(32\pi)=E_p \cdot (\tan \delta (p) / p )$ at $m_\pi / m_\rho = 0.491$ and $0.593$
obtained by substituting our data for $\Delta W_p$ into (\ref{Lusher.eq}).
In order to obtain the phase shift
for various momenta at the physical pion mass, 
we extrapolate our data
with the following fitting assumption :
$T/(32\pi)= A_{10} \cdot (m_\pi^2) + A_{20} \cdot (m_\pi^2)^2
+ A_{01} \cdot (p^2) + A_{02} \cdot (p^2)^2         
+ A_{11} \cdot (m_\pi^2 p^2)$.
The fit curves are also plotted in Fig.~\ref{Fig3.fig}.

\begin{figure}[t]
\vspace*{-0.8cm}
\centerline{\epsfxsize=7.0cm \epsfbox{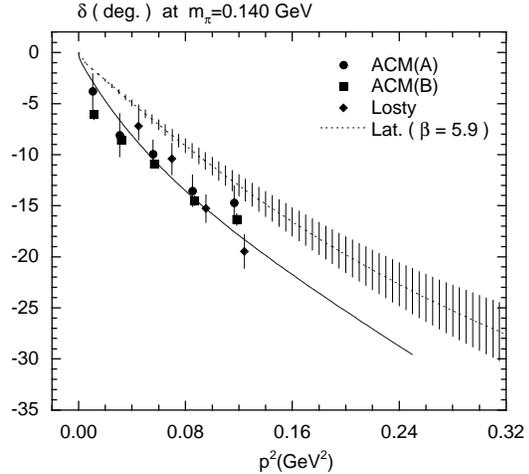}}
\vspace*{-1.0cm}
\caption{\label{Fig4.fig}
Comparison of our results of scattering phase shift $\delta(p)$ and 
experiments.
}
\vspace*{-0.6cm}
\end{figure}

In Fig.~\ref{Fig4.fig}
we compare the calculated phase shift $\delta(p)$
at physical pion mass with the fit above 
with experiments~\cite{PHSH_expt}.
Our results for $\delta (p)$ are 30\% smaller in magnitude than those of 
experiments,  
and our result for scattering length, 
$a_0 m_\pi = -0.0272(16)$, 
differs from the ChPT prediction given by  
$a_0 m_\pi = -0.0444(10)$~\cite{Gasser-Leutwyler:Bijinens}, but
consistent with the result of JLQCD at the same lattice spacing, 
$a_0 m_\pi = -0.0300(31)$~\cite{JLQCD}.

A possible cause of the discrepancy is finite lattice spacing effects. 
The JLQCD results for scattering length show sizable scaling 
violation~\cite{JLQCD}. 
Hence that of the scattering phase shift cannot be considered small.
Further calculations nearer to the continuum limit or calculations with 
improved actions are desirable.
%
%
%
%
\hfill\break

This work is supported in part by Grants-in-Aid of the Ministry of Education 
(Nos.
11640294, 
12304011, 
12640253, 
12740133, 
13640259, 
13640260, 
13135204, 
14046202, 
14740173, 
). 
VL is supported by the Research for Future Program of JSPS
(No. JSPS-RFTF 97P01102).
Simulations were performed on the parallel computer CP-PACS. 
%
%

%
%
\end{document}